\documentclass[twocolumn,aps,preprintnumbers,amsmath,amssymb,superscriptaddress,nofootinbib]{revtex4}

\usepackage{slashed}
\usepackage{epsfig,latexsym,cancel,amssymb,amsmath,verbatim,mathrsfs}
\usepackage{color}
\usepackage{graphicx,subfigure,multirow}

\usepackage{dcolumn}
\usepackage{bm}
\usepackage[all]{xy}
\usepackage{color}
\usepackage[usenames,dvipsnames]{xcolor}
\usepackage[unicode=true,pdfusetitle,
 bookmarks=true,bookmarksnumbered=false,bookmarksopen=false,citecolor=Turquoise,
 breaklinks=false,pdfborder={0 0 1},backref=false,colorlinks=true,pdfpagemode=FullScreen]
 {hyperref}
\usepackage{amssymb}
 \usepackage{longtable}
\usepackage{epsfig}
\usepackage{amsmath}
\usepackage{pstricks}
\usepackage{bbold}
\usepackage{slashed}
\usepackage{amssymb}
\usepackage{graphicx}
\usepackage{hyperref}
\usepackage{verbatim}
\usepackage{multirow}
\usepackage{chemarrow}
\usepackage{hhline}
\makeatletter

\newcommand{\beq}{\begin{equation}}
\newcommand{\eeq}{\end{equation}}
\newcommand{\bea}{\begin{eqnarray}}
\newcommand{\eea}{\end{eqnarray}}

\begin{document}

\preprint{
\begin{minipage}[b]{1\linewidth}
\begin{flushright}
IPMU16-0141 \\
KEK-TH-1935
 \end{flushright}
\end{minipage}
}

\title{Surviving scenario of stop decays for ATLAS $\ell+jets+\slashed{E}^{miss}_T$ search}

\author{Chengcheng Han}
\email[]{chengcheng.han@ipmu.jp }
\affiliation{Kavli IPMU (WPI), The University of Tokyo, Kashiwa, Chiba 277-8583, Japan}
\author{Mihoko M.\,Nojiri}
\email[]{nojiri@post.kek.jp }
\affiliation{Kavli IPMU (WPI), The University of Tokyo, Kashiwa, Chiba 277-8583, Japan}
\affiliation{KEK Theory Center, IPNS, KEK, Tsukuba, Ibaraki 305-0801, Japan}
\affiliation{The Graduate University of Advanced Studies (Sokendai),Tsukuba, Ibaraki 305-0801, Japan}
\author{Michihisa Takeuchi}
\email[]{michihisa.takeuchi@ipmu.jp}
\affiliation{Kavli IPMU (WPI), The University of Tokyo, Kashiwa, Chiba 277-8583, Japan}
\author{Tsutomu T. Yanagida}
\email[]{tsutomu.tyanagida@ipmu.jp}
\affiliation{Kavli IPMU (WPI), The University of Tokyo, Kashiwa, Chiba 277-8583, Japan}


\begin{abstract}
Recently ATLAS reported a $3.3\sigma$ excess in the stop search with $\ell+jets+\slashed{E}_T^{miss}$ channel. 
We try to interpret the signal by a light stop pair production in the MSSM. We find: (1) simple models where stop 
decays into a higgsino or a bino are not favored. (2) an extension of them 
can explain the data at $2\sigma$ level without conflicting with the other search channels.
A surviving possibility includes a light stop and a light higgsino, which is expected in a natural SUSY scenario.
\end{abstract}

\maketitle
\section{Introduction.}
Supersymmetry (SUSY) is one of the most fascinating models beyond the Standard Model (SM),
which can simultaneously solve the naturalness problem, explain the cosmic dark matter and achieve the gauge coupling
unification. Searching for the SUSY particles is an important task for LHC especially after the discovery of Higgs.
With the recent 13~fb$^{-1}$ dataset of LHC,  a gluino and a stop lighter than 1.8 TeV  and 900 GeV have been excluded out respectively if the mass splitting between the stop and the LSP is large enough, and the mass of an electroweakino has been excluded to 0.4-1 TeV depending on the decay mode. All the results push SUSY scale much higher than before, which further challenges our understanding on the naturalness problem.   

Among those tens of super-particles, the top partner-stop,  plays the most important role to understand the naturalness in SUSY models. 
In the framework of natural supersymmetry, a light stop and higgsino are usually predicted~\cite{naturalSUSY}. Lots of works have been done on searching for such light stops and higgsinos~\cite{stop}. Recently ATLAS reported some excesses in the stop search with $\ell+jets+\slashed{E}^{miss}_T$ channel~\cite{ATLAS:2016ljb}. Seven signal regions are defined and three of them observed more than $2\sigma$ excesses, among which the most significant one could reach $3.3\sigma$. Although no significant excess is reported by CMS in the same channel~\cite{CMS:2016vew},  
they observe mild excesses in several signal regions. Even in the $0$ lepton channel searches~\cite{CMS:2016hxa,CMS:2016inz,ATLAS:2016jaa}, 
ATLAS and CMS both show small excesses with a lower significance. If they are true signals it is likely due to new particles. 

In this paper, we study the interpretation of the excesses as light stop pair production in the minimal supersymmetric standard model~(MSSM). 
Although ATLAS and CMS have done the similar analyses based on the simplified models, the results might be changed in a concrete model of MSSM.
We show that the simplified models where stop decays into a higgsino or a bino are not favored by other stop searches while extended models could provide a better fit to the excess.

This paper is organized as follows. We first briefly overview the ATLAS $\ell+jets+\slashed{E}^{miss}_T$ search and the other constraints at the LHC
in Section~\ref{sec:stop}. In Section~\ref{sec:model} we interpret the excess by a light stop pair production in the MSSM. 
In Section~\ref{sec:darkmatter}, we discuss the dark matter constraints, such as 
the relic abundance and the direct detection constraints.
We draw our conclusions in Section \ref{sec:conclusion}.

\section{stop searches}
\label{sec:stop}
\subsection{ATLAS $	\ell+jets+\slashed{E}^{miss}_T$ search}
The ATLAS $\ell+jets+\slashed{E}^{miss}_T$ search mainly aims at stop pair production followed by stop decay into top and neutralino, where one of the resulting top decay leptonically.
Although more details are found in~\cite{ATLAS:2016ljb} we summarize some of the selection cuts and the results in Table~\ref{tab:atlas}. 
Seven signal regions are defined for the searches.
Among them, the signal regions of {\tt DM\_low}, {\tt bC2x\_diag} and {\tt SR1} observe $3.3\sigma$, $2.6\sigma$, $2.2\sigma$ excesses, respectively. 
Interestingly, the {\tt SR1} signal region was already defined before and 
an excess around $2.3\sigma$ had been reported even with the 3.2~fb$^{-1}$ data~\cite{Aaboud:2016lwz}. 

\begin{table*}[t!]
\centering
\begin{tabular}{c|c|c|c|c|c|c|c}
\hline Signal  region & SR1  & tN\_high  & bC2x\_diag & bC2x\_med & bCbv & DM\_low  & ~~~DM\_high~~~ \\ 
\hline\hline 
$(n_j,n_b)$ & $(\ge 4, \ge 1)$ & $(\ge 4, \ge 1)$ & $(\ge 4, \ge 2)$ & $(\ge 4, \ge 2)$ & $(\ge 2,  = 0)$  & $(\ge 4, \ge1)$  & $(\ge 4, \ge1)$  \\ 
$E\!\!\!/_T$ [GeV] & 260 & 450 & 230 & 210 & 360 & 300 & 330 \\ 
$m_T$ [GeV] & 170 & 210 & 170 & 140 & 200 & 120 & 220 \\
$am_{T2}$ [GeV] & 175 & 175 & 170 & 210 & - & 140 & 170 \\
\hline  \hline Total background    & $24\pm3$     &$3.8\pm 0.8$     &$22\pm3$      &$13\pm2$      &$7.8\pm1.8$       &$17\pm2$        &$15\pm2$  \\
\hline  Observed               &37   &5   &37   &14   &7   &35   &21 \\
\hline \hline $p_0(\sigma)$    & 0.012(2.2) &0.26(0.6) & 0.004(2.6) & 0.40(0.3) & 0.50(0) & 0.0004(3.3) &0.09(1.3) \\
\hline $N^{\rm limit}_{\rm obs.}$(95\% CL) &26.0 &7.2 & 27.5 & 9.9 & 7.2 & 28.3 & 15.6 \\
\hline
\end{tabular}
\caption{Summary of some of the selection cuts and the results of the seven signal regions defined in
ATLAS stop search in $\ell+jets+\slashed{E}^{miss}_T$ channel.}
\label{tab:atlas}
\end{table*}%

We provide brief comments on the search results.
\begin{itemize}
\item[(1)]  These seven signal regions are not exclusive. {\tt SR1}, {\tt bC2x\_diag} and {\tt DM\_low}, where the excesses are observed, 
could share the same signal events.
\item[(2)]  The {\tt DM\_low}  and {\tt DM\_high} applies similar cuts except a lower $m_T$ cut and a tighter $\Delta \phi(\slashed{E}_T^{miss}, j_{i})$ cut in {\tt DM\_low}. Since no excess is observed in the {\tt DM\_high}, hard $m_T$ events are not preferred.
\item[(3)]  The {\tt bC2x\_med} requires very high $p_T$ bottoms 
compared with the {\tt bC2x\_diag} . Since no excess is observed in the {\tt bC2x\_med}, presence of hard bottoms is not preferred. 
\end{itemize}

In the following we focus on the excess 
in the signal region {\tt DM\_low}.
The estimated event number of background is $17\pm2$ 
while the observed number of events is 35 in {\tt DM\_low}. From the background+signal hypothesis, 
estimated $2\sigma$-confidence interval for the number of signal is [7.4, 32.6] and the central value is 18.0.
We summarize the confidence intervals used in our analysis in Table~\ref{tab:signal}. 
All the other signal regions listed in the Table~\ref{tab:atlas} are considered for setting 95\% C.L. exclusion contours. 
Note that we don't combine the results from different signal regions as they are not statistically independent. 
Instead, we overlay the preferred signal regions and the exclusion contours in the following analysis. 

\begin{table}[h!]
\centering
\begin{tabular}{|c|c|c|c|c|c|c|c|}
\hline Signal region  & $2\sigma$ upper &$1\sigma$ upper& central&$1\sigma$ lower& $2\sigma$ lower \\
\hline DM\_low   & 32.6 & 24.7 & 18.0 & 12.2 & 7.4 \\
\hline
\end{tabular}
\caption{Confidence-intervals of the number of signal in {\tt DM\_low} used in our analysis.}
\label{tab:signal}
\end{table}%

\subsection{Other stop search constraints}
There are other stop searches 
based on 0 lepton, 1 lepton, and 2 leptons both from ATLAS and CMS, which constrain the stop parameter space.
We have found the 1 lepton search constraints from CMS~\cite{CMS:2016vew} gives more or less similar to that of the ATLAS,
therefore, considering the signal regions with no excess in ATLAS 1 lepton search is enough.
The two lepton search channels do not set strong constraints~\cite{ATLAS:2016xcm}, and we did not consider them in this paper.
The CMS 0 lepton search channels set slightly stronger bound than those of ATLAS, 
therefore, we consider the following two hadronic stop search channels at CMS to set the 95\% CL exclusion.
The CMS boosted top search requires top tagging and is especially sensitive to 
the stop with a large mass splitting with LSP, where boosted tops are expected in the final state. 
The CMS hadronic stop search aims at more conventional topologies from stop decay. 
In the following, the ATLAS 1 lepton search is denoted as {\tt ATLAS\_1L}, 
the CMS hadronic stop search~\cite{CMS:2016hxa} is denoted as {\tt CMS\_hadronic}
and CMS boosted top search~\cite{CMS:2016inz} is denoted as {\tt CMS\_boosted}.

\section{Simplified models}
\label{sec:model}
In this paper we interpret the excess from a light stop pair production and their decays.
We assume the lightest stop dominantly consists of the right handed stop~($\tilde{t}_R$).
The $\tilde{t}_L$ usually is accompanied by $\tilde{b}_L$ which is close in mass, 
and it is constrained strongly with a bound up to about 1 TeV~\cite{CMS:2b}.
We first consider a simple 1-step decay from the right-handed stops.  
Depending on the dominant component of the LSP, the constraints are modified as the decay modes change.
We scan the parameter space to figure out the corresponding $2\sigma$ favored region in {\tt DM\_low}.

Our simulation for {\tt ATLAS\_1L } and {\tt CMS\_hadronic} searches are based on {\tt MadGraph + Pythia}~\cite{Alwall:2014hca,Sjostrand:2006za}, 
and the generated signal events are passed to {\tt Delphes3}~\cite{Ovyn:2009tx} for a detector simulation.
For recasting {\tt CMS\_boosted} search 
we simply compare the cross sections rescaled with $BR(\tilde{t} \to t \tilde{\chi})^2$ with 
the upper bound of the cross section given in the CMS paper 
assuming the best sensitivity is from two boosted top tagged events.

\begin{figure}[h]
 \includegraphics[width=0.6\textwidth]{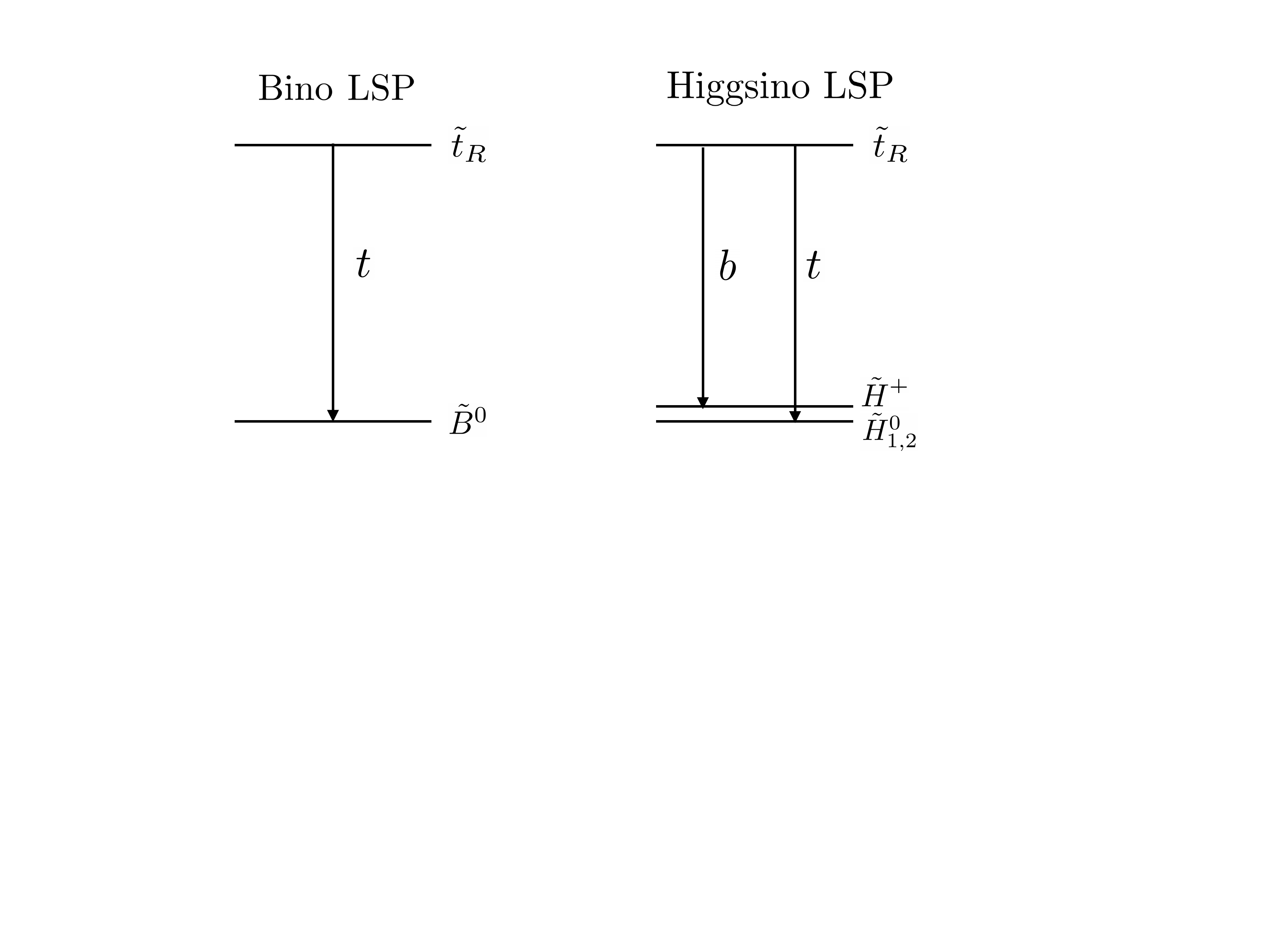} 
 \vspace{-4.0cm}
   \caption{Stop decay in the simplified model.  } 
\label{decay_bino}
\end{figure}

\subsection{Bino LSP}
\begin{figure}[t]
\includegraphics[width=0.4\textwidth]{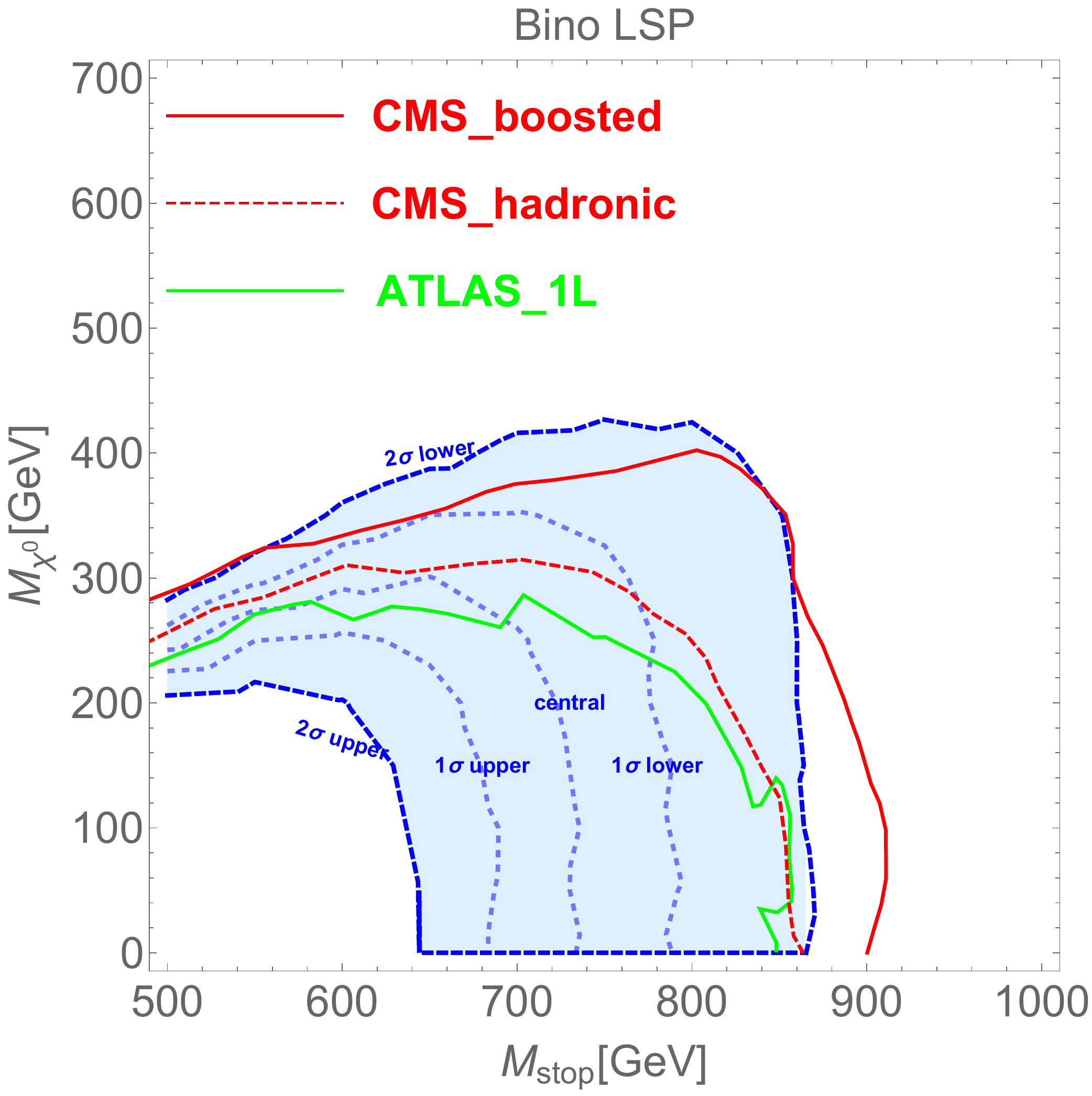}
\caption{$2\sigma$ favored region and the excluded region from the Bino LSP model.  } 
\label{bino}
\end{figure}

We first consider the simplest model: the lightest supersymmetric particle~(LSP) is bino and the next-lightest supersymmetric particle~(NLSP) is $\tilde{t}_R$. As shown in Fig~\ref{decay_bino}, the only decay channel is $\tilde{t}\rightarrow \tilde{\chi}^0_1+ t$ in this model, which is frequently assumed for many analyses in ATLAS and CMS.
 In Fig~\ref{bino} we show the $2\sigma$ favored region by the excess in {\tt DM\_low} 
 and the 95\% C.L. exclusion contours on the $(m_{\tilde t}, m_{\chi^0_1})$ plane
 from {\tt ATLAS\_1L}, {\tt CMS\_hadronic}, and {\tt CMS\_boosted} searches.
The strongest limit is from {\tt CMS\_boosted} search
because the tops in the final state tend to be boosted.
We find all the $1\sigma$ favored region are excluded out by {\tt CMS\_boosted} and
almost all the $2\sigma$-region is excluded.

\subsection{Higgsino LSP}
\begin{figure}[t]
\includegraphics[width=0.4\textwidth]{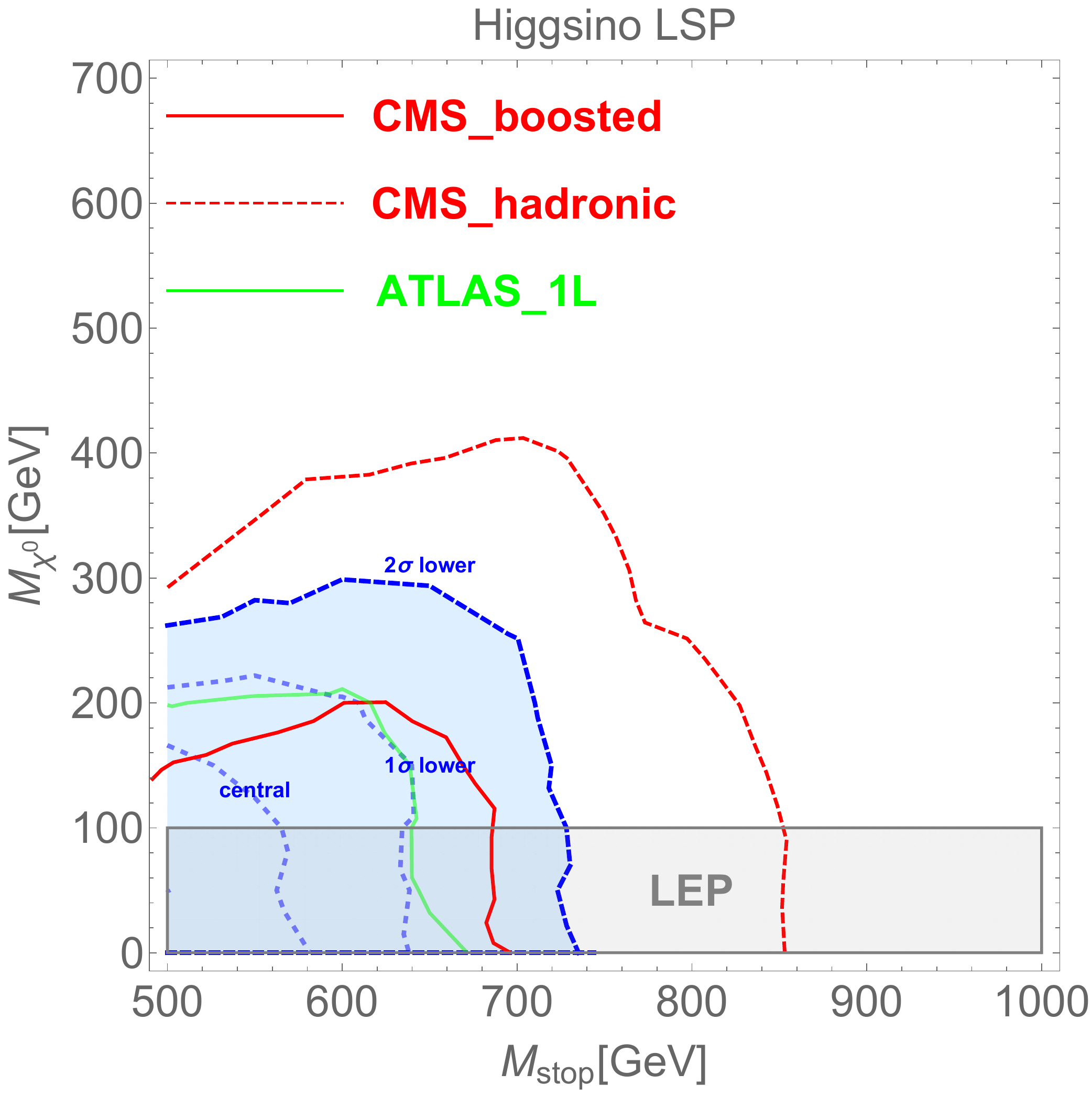}
\caption{$2\sigma$ favored region and the excluded region from the Higgsino LSP model.   } 
\label{higgsino}
\end{figure}

To ease the tension from {\tt CMS\_boosted} stop search, we may consider a model where the stop has 
other decay channels than the direct decay into top and LSP.
As an example, we consider another simplified model where LSP is higgsino.
For the higgsino LSP case, the higgsinos ($\chi^0_{1,2}$) and charged higgsino ($\chi_1^\pm$) is naturally degenerate.
As the typical mass difference is $\sim$ GeV, 
we cannot observe the decay products from those particles essentially
and all particles behaves like LSP in terms of the collider signature at the LHC.
The branching ratios are 
$BR({\tilde t}_R \to t \chi^0_{1,2}) \sim BR({\tilde t}_R \to b \chi^+_1) \sim 50\%$ 
when the phase space suppression due to the top mass is negligible. Thus, the contribution of the two boosted tops is reduced by a factor $1/4$.
Since {\tt CMS\_boosted} stop search relays on two boosted tops in the final states, 
the constraint becomes much weaker. On the other hand, the $1 \ell$ signal 
can originate from events with one stop decay into a top and the other into a bottom, thus, 
there are no strong reduction factor.
In Fig~\ref{higgsino} we show the signal preferred region and several exclusion contours. In the plot we fixed $\tan \beta=10$.  
Note the branching ratios of stop decay do not change a lot when we vary the $\tan\beta$ because higgsino-stop coupling
only comes from the yukawa. The gray region is limited by direct chargino search from LEP\cite{Abbiendi:2002vz}.
The {\tt CMS\_boosted} stop search becomes significantly weaker, and there appears 
a large region not excluded by either {\tt ATLAS\_1L} and {\tt CMS\_boosted} 
but in the $2\sigma$ favored region.
However, the {\tt CMS\_hadronic} constraints are still strong enough to exclude the whole $2\sigma$ signal favored region. 
It is because reducing $BR({\tilde t}_R \to t \chi^0_{1,2})$ also reduce 
the number of 1 lepton signals, while the conventional 0 lepton signals are not reduced and 
it results in a similar sensitivity to the Bino LSP case.

\subsection{Higgsino + Bino LSP}
Although by reducing $BR({\tilde t}_R \to t \chi^0_1)$ the tension between {\tt CMS\_boosted} and {\tt ATLAS\_1L} searches could be eased,
it also reduces the signal events and makes the conventional hadronic stop search relatively more effective. 
To avoid this situation, keeping more signal events while reducing top branching ratio is necessary, therefore,
it is preferable to find a way to make the $BR({\tilde t}_R \to b \chi^+)$ also contribute to enhance the lepton signals.
We consider here a model where NLSP is higgsino and LSP is bino. 
Since the chargino will decay into $W$ plus neutralino, 
one lepton could come from the $W$ decay through a two step decay.
We set the mass difference between stop and higgsino 150~GeV to forbid $t + {\tilde H}^0_{1,2}$ decay
to make the discussion simpler.
We have checked that opening $t + \tilde{H}^0_{1,2}$ mode also gives similar final results.

\begin{figure}[h]
 \includegraphics[width=0.7\textwidth]{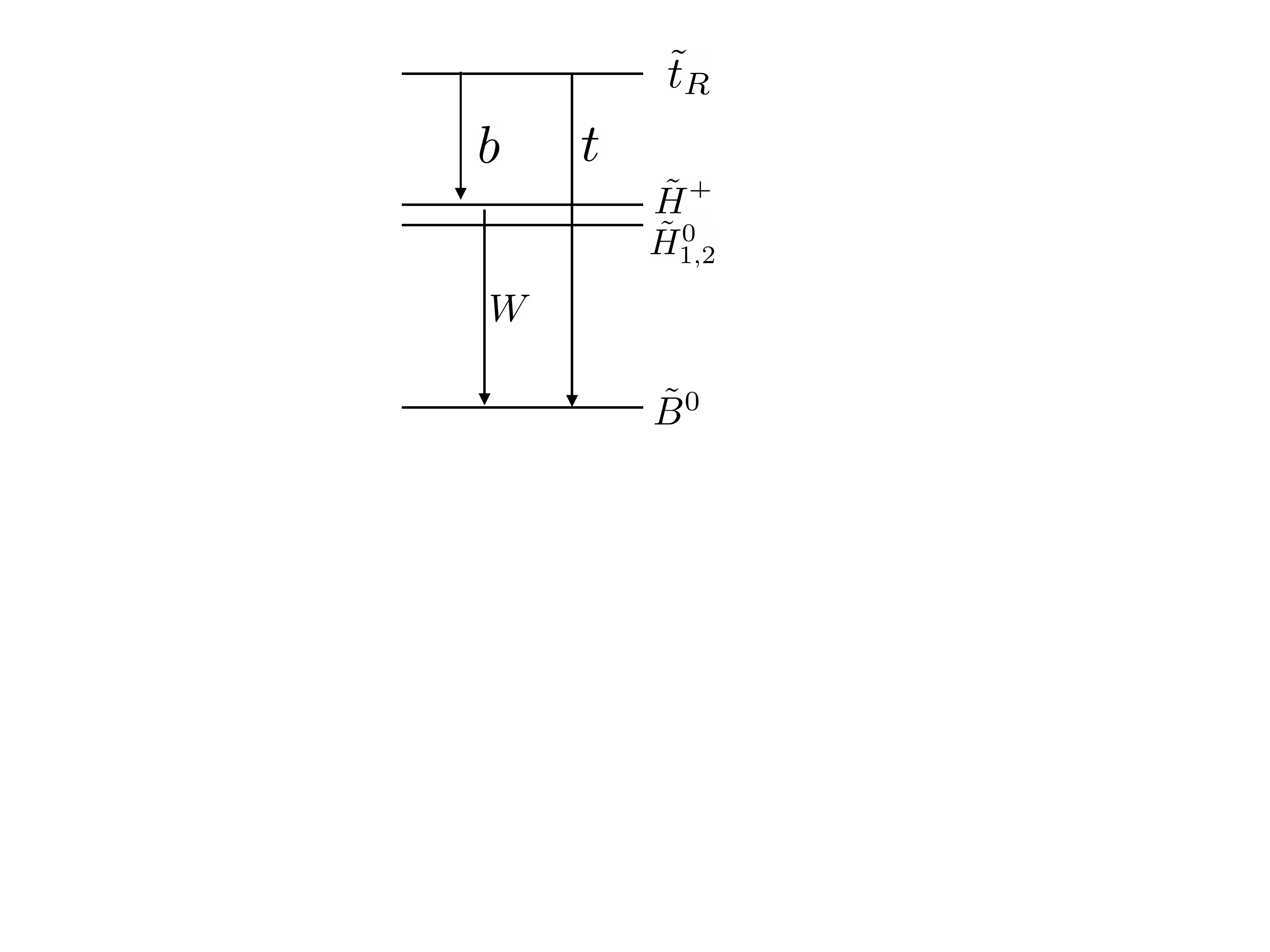}
 \vspace{-5.0cm}
   \caption{Stop decay in the Higgsino + Bino LSP model. } 
\label{decay_higgsino_bino}
\end{figure}

\begin{figure}[h]
\includegraphics[width=0.4\textwidth]{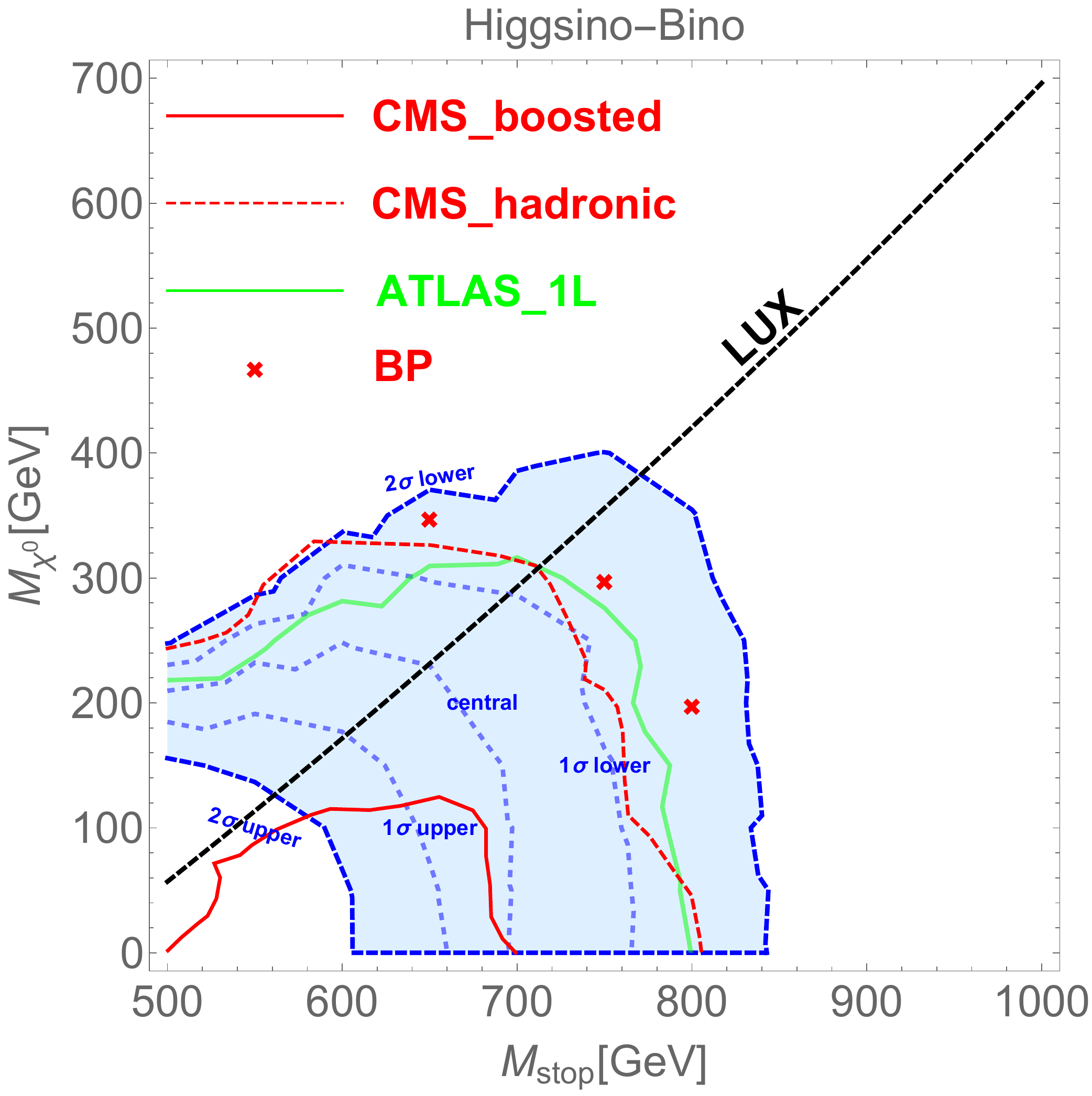}
\caption{$2\sigma$ favored region and the excluded region in Higgsino+Bino LSP model. 
We fix $m_{\tilde{t}}-m_{\tilde{\chi}^{\pm}}=150$ GeV and $\tan\beta=10$.  
Left side of the LUX line has been excluded by the dark matter direct detection experiments.} 
\label{fig:higgsinobino}
\end{figure}

In Fig~\ref{fig:higgsinobino}, we show our simulation results recasting the {\tt ATLAS\_1L}, {\tt CMS\_hadronic}, 
and {\tt CMS\_boosted} searches on the $(m_{\tilde t}, m_{\chi^0_1})$ plane assuming $m_{\chi^+_1} = m_{\tilde t} - 150$~GeV.
Large parameter region in the Higgsino + Bino LSP model satisfying 
$m_{\tilde t} \sim 800$~GeV with $m_{\chi^0_1} \lesssim 350$~GeV
or $650$~GeV$ \lesssim m_{\tilde t} \lesssim 800$~GeV with $m_{\chi^0_1}\sim 350$~GeV
is found consistent within $2\sigma$ to all constraints
although $1\sigma$ favored region is still excluded by both {\tt ATLAS\_1L} and {\tt CMS\_hadronic}.

{\tt ATLAS\_1L} also provides the $\slashed{E}_T^{miss}$ and $m_T$ distributions in {\tt DM\_low}.
We have selected three benchmark points, which are indicated with crosses in Fig~\ref{fig:higgsinobino}, and show the expected distributions.
The background distributions we just take from the ATLAS plots.
The benchmark points with stop-bino mass 650-350 GeV, 750-300 GeV, 800-200 GeV
predict the signal events in {\tt DM\_low} to be 9.4 
($1.6\sigma$), 9.8 ($1.5\sigma$) and 8.3 ($1.8\sigma$), respectively. 
The numbers in the parentheses indicate
the statistical deviations assuming the corresponding signal injection, 
to be compared with $3.3~\sigma$ of no signal assumption.
Although all the benchmark points are consistent within $2 \sigma$ based on 
the total number of events in {\tt DM\_low} the predicted distributions are different.
We find compressed spectrum is slightly preferred as the overflow bin doesn't contain much events.
In the future these distributions would be important to distinguish between models.

 \begin{figure}[t]
 \includegraphics[width=0.4\textwidth]{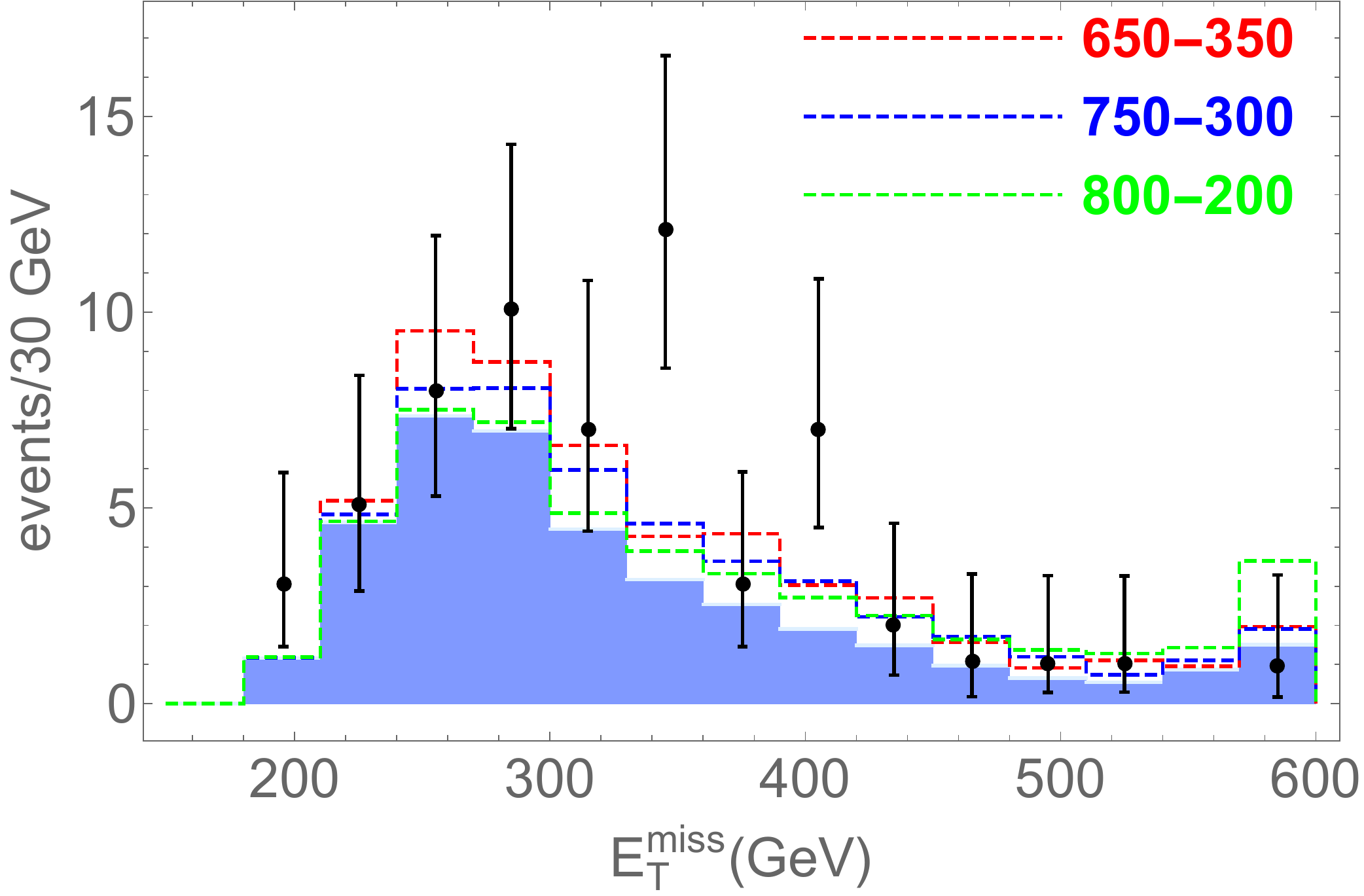}
  \includegraphics[width=0.4\textwidth]{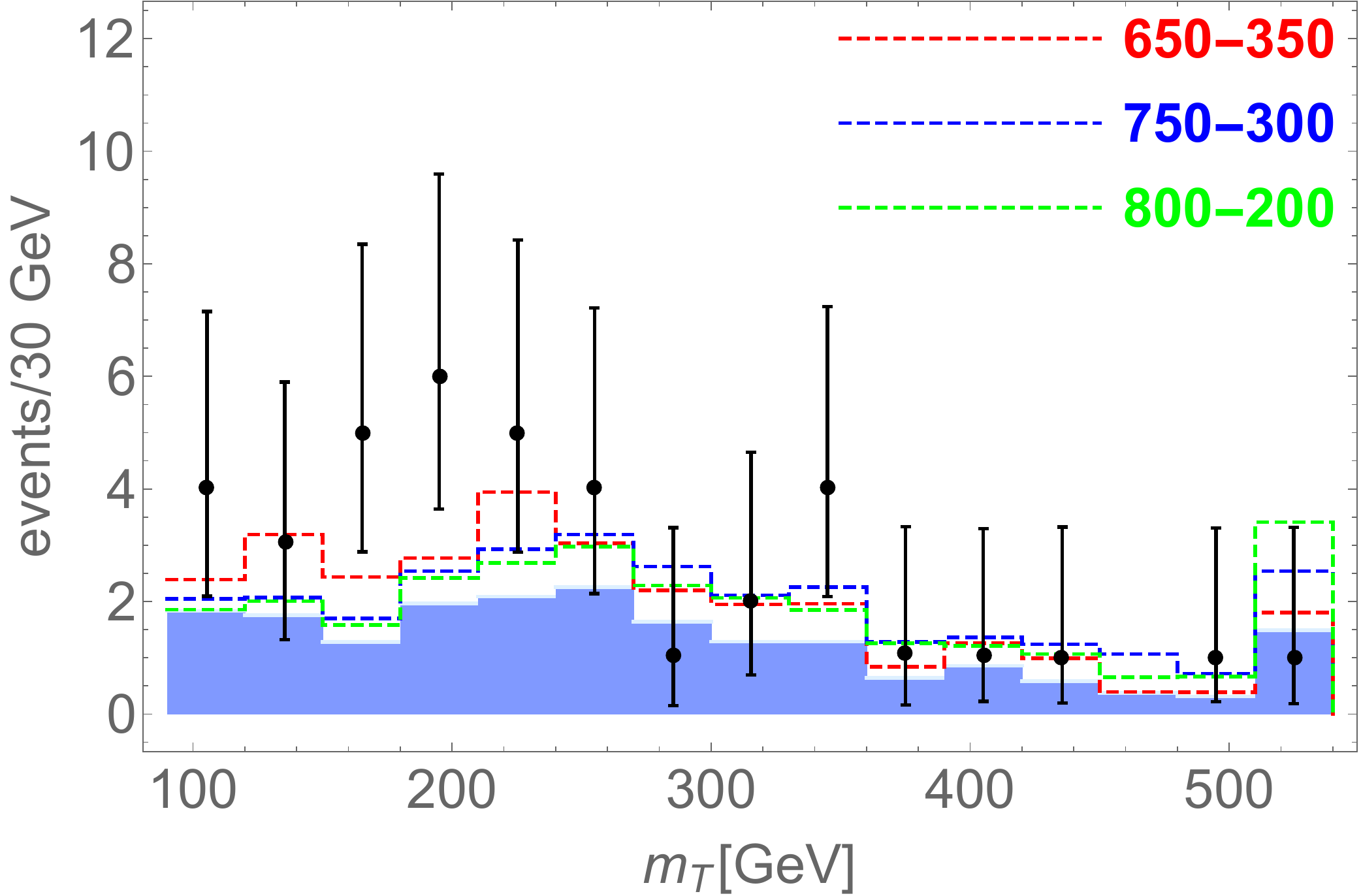}
\caption{$\slashed{E}^{miss}_T$ and $m_T$ distribution for the three benchmark points 
in the consistent region of the Higgsino + Bino LSP model. 
The numbers in legend are the stop and the LSP masses. The right most bin contains overflow events.} 
\label{distributions}
\end{figure}

Although we only consider the Higgsino-Bino case (we denote the case with Higgsino NLSP and Bino LSP as Higgsino-Bino, etc. in the following), 
we would like give some comments  on other possibilities. 
Since the coupling between stop~($\tilde{t}_R$) and Wino is suppressed by the neutralino mixing,
the Wino-Bino (Wino-Higgsino) cases are essentially reduced to the Bino (Higgsino) LSP case.
With the large L-R mixing in stop sector 
we can tune the relative branching ratios by the stop mixing angle 
and the collider signature of the Wino-Bino case could be similar to the Higgsino-Bino model.
Three remaining possibilities are  Bino-Higgsino, Bino-Wino, and Higgsino-Wino cases. 
For the Bino-Higgsino case, 
the stop dominantly decays into a higgsino,
as stop couples higgsinos thorough the top yukawa coupling,
which is much stronger than bino through the gauge coupling, 
and due to the phase space suppression, therefore, it is again reduced to the Higgsino LSP case. 
The Higgsino-Wino case is similar to the Higgsino-Bino case but the higgsino decay pattern is more complex,
and branching ratio into $W$ would be suppressed.
The Bino-Wino case might be an interesting possibility but constrained by the multi-lepton searches.
On the other hand, direct detection constraint must be significantly weak for this case
whereas indirect searches might be dangerous.
Also notice that the long lived charged particle searches at 8 TeV~LHC already 
set the lower bound of the wino mass around 270 GeV~\cite{Aad:2013yna}, 
and the allowed parameter space would be highly constrained. 

\section{Dark matter}
\label{sec:darkmatter}
In the previous section, we study the case where bino is the LSP.
It might overclose the universe because the pair annihilation cross section of bino like dark 
matter is rather small. 
An interesting possibility is to resort to the slepton co-annihilation. 
If a slepton is nearly degenerate with the bino LSP $\sim 5$~GeV~\cite{Citron:2012fg}
it would help to reduce the relic abundance to the observed one. 
One may wonder inserting slepton in the spectrum will affect the decay mode of the chargino.
As the coupling between slepton and higgsino is proportional to the corresponding lepton yukawa coupling
only a light stau with a large $\tan\beta$ could affect the branching ratios.
Note that the chargino decaying into neutralino plus $W$ is 
dominated by the longitudinal mode of $W$ (from $H$-$\tilde{H}$-$\tilde{B}$ interaction), and 
not suppressed by the Higgsino-bino mixing. 
The stau decaying channel only becomes sizable when the $y_\tau$ is enhanced 
by very large $\tan\beta$. Taking $\tan\beta=10$ as an example, we find in our $2\sigma$ favored region, 
the stau decay mode at most has a branching ratio of $\sim 12\%$. 
If we reduce the $\tan\beta=5$, this decay branching ratio only takes 3\%, so our Fig~\ref{fig:higgsinobino} is essentially unchanged.
Moreover, the $\tau$ emitted 
from the ${\tilde \tau}$ decay is not detectable at the collider
due to the small mass difference.

The dark matter direct search also constrains the Higgsino-Bino model 
due to the mixing between higgsino and bino. In Fig~\ref{fig:higgsinobino} we draw 
the sensitivity of the latest LUX results~\cite{Akerib:2016lao}
the parameter space on the left side of the line has been excluded. 
In the plot we fix $m_{\tilde{t}}-m_{\tilde{\chi}^{\pm}}=150$ GeV and $\tan\beta=10$, sign$(M_1/\mu)=1$, and the Higgs mass is also tuned to be 125 GeV.  
We find one of our benchmark points has been excluded out and one just lives the boundary of exclusion line.
Therefore the future dark matter direct search will be important to test this scenario if the signal is confirmed at the LHC. 
Note that in the case of sign$(M_1/\mu)=-1$, there exists a blind spot region for dark matter direct search
when $M_1+\mu \sin 2\beta=0$ is satisfied\cite{Cheung:2012qy, Huang:2014xua}.

We may also assume R-parity is violated and bino could decay into SM particles 
with the life time long enough in a collider time scale. In such a case, 
the constraints discussed in this section are not applicable.

\section{Conclusion}
\label{sec:conclusion}
In this paper, we try to interpret the $3.3\sigma$ excess reported in ATLAS $\ell+jets+\slashed{E}_T^{miss}$ channel 
by a light stop pair production in the MSSM. 
After considering the consistency with other stop searches, we find: (1) simple models where stop 
decays into a higgsino or a bino are not favored by the other search channels 
(2) an extension of them, Higgsino-Bino LSP,  can explain the data at $2\sigma$ level without conflicting with 
the other search channels. It is remarkable that the SUSY spectrum in the surviving stop 
decay scenario for the possible excess, that is a light stop and a light Higgsino, is nothing 
but what we expect in a natural SUSY.

\noindent {\bf{Acknowledgements}}
The authors wish to thank Dr. Lei Wu for valuable discussions and for the technical helps.
CCH, MT, MMN and TTY are supported by World Premier International Research Center Initiative (WPI Initiative), MEXT, Japan.
MMN is also supported by Grant-in-Aid for Scientific research Nos.\, JP16H06492, JP16H03991 and JP26287039.
MT is supported by Grant-in-Aid for Scientific Research Numbers JP16H03991 and JP16H02176. 
TTY  is supported by JSPS Grants-in-Aid for Scientific Research Nos. JP26287039, JP26104009  and JP16H02176.

\end{document}